%% file: main.tex
\documentclass[twocolumn]{revtex4-2}
\usepackage{chemformula} 
\usepackage{verbatim}
\usepackage{graphicx}
\usepackage{grffile}
\usepackage{xcolor}
\usepackage{soul}
\usepackage{hyperref}
\graphicspath{{mainplots/}} 

\newcommand*\vc[1]{\mathbf{#1}}
\newcommand*\tx[1]{\mathrm{#1}}
\newcommand*\wn{cm$^{-1}$}

\usepackage[symbol]{footmisc}

\newcommand{\1}[1]{#1}

\begin{document}
\title[Protein-Water Energy Transfer] {Protein-Water Energy Transfer via Anharmonic Low-Frequency Vibrations}

\author{Brandon Neff}
\author{Matthias Heyden}
\email{mheyden1@asu.edu}
\affiliation{School of Molecular Sciences, Arizona State University, Tempe, AZ 85287, U.S.A.}

\date{\today}

\begin{abstract}
\normalsize
Heat dissipation is ubiquitous in living systems, which constantly convert distinct forms of energy into each other.
The transport of thermal energy in liquids and even within proteins is well understood but kinetic energy transfer across a heterogeneous molecular boundary provides additional challenges.
Here, we use atomistic molecular dynamics simulations under steady-state conditions to analyze how a protein dissipates surplus thermal energy into the surrounding solvent.
We specifically focus on collective degrees of freedom that govern the dynamics of the system from the diffusive regime to mid-infrared frequencies.
Using a fully anharmonic analysis of molecular vibrations, we analyzed their vibrational spectra, temperatures, and heat transport efficiencies.
We find that the most efficient energy transfer mechanisms are associated with solvent-mediated friction.
However, this mechanism only applies to a small number of degrees of freedom of a protein.
Instead, less efficient vibrational energy transfer in the far-infrared dominates heat transfer overall due to a large number of vibrations in this frequency range. 
A notable by-product of this work is a highly sensitive measure of deviations from energy equi-partition in equilibrium systems, which can be used to analyze non-ergodic properties.
\end{abstract}

\maketitle




\input{introduction.tex}
\input{theory.tex}
\input{methods.tex}

\input{results_and_disc.tex}

\input{conclusion.tex}

\raggedbottom

\begin{acknowledgements}
This work is supported by the National Institute of General Medical Sciences (R01GM148622). The authors acknowledge Research Computing at Arizona State University for providing high performance computing resources~\cite{sol} used in this work.
\end{acknowledgements}

\newpage
\input{main.bbl}

%
%
\input{suppinfo.tex}


\end{document}

%% file: introduction.tex
\section{Introduction}
The kinetic energy of a classical system in equilibrium is distributed equally among all independent degrees of freedom, resulting in a state of maximum entropy.
However, many systems of interest, especially living systems in biology and many nano-scale devices and circuits,\cite{pop2010energy} are not in equilibrium, and thus do not adhere to this rule.
A hallmark of life, among others, is conversion of energy.\cite{yewdall2018hallmarks}
Light absorption or consumption of energy-rich nutrients provides energy inputs used to generate ATP, build up other energy-rich compounds, store energy in concentration gradients across lipid membranes, {\em etc}.
Both storage and eventual release of energy to drive critical biochemical processes involve energy conversion steps that characterize the metabolism of a given species. 
A necessary byproduct of such energy conversions is heat dissipated into the environment.

Macroscopically, {\em e.g.}, especially for warm-blooded land animals and birds, the interface with the surrounding atmosphere generates a bottleneck for energy dissipation, and the resulting thermogenesis plays a critical role for temperature regulation.\cite{grabek2024evolution,morrison2008central} 
Microscopically, heat generated by biochemical processes in solution dissipates via microconvection~\cite{howard2019cytoplasmic} and acoustic modes that propagate through the respective medium at the speed of sound~\cite{milhamont2024phonon}.
In related contexts, phonon-induced friction has been described for molecules non-covalently bound to surfaces.\cite{farahvash2024theory}
In water, viscoelastic effects result in fast propagation velocities of acoustic modes (akin to amorphous or even crystalline ice) on short time and length scales (picoseconds, nanometers) that transition to the macroscopic sound velocity of the liquid in a temperature-dependent way.\cite{sciortino1994sound,capponi2018structural}
Heat transfer in bulk liquids, homogeneous condensed matter systems and even proteins have been studied extensively.\cite{leitner2008energy}
However, the microscopic details of heat transfer at inhomogeneous surfaces, such as biomolecules and their solvation environment, are less well understood.\cite{noneq_small_sys}

Early femtosecond infrared (IR) spectroscopy experiments have indicated two separate modes of solute-solvent energy transfer: a fast (7–8~ps) and a slow (20~ps) component.\cite{lian1994energy}
Similar kinetics were obtained for vibrational decay and temperature relaxation upon light-induced dissociation of ligands from the heme-group of myoglobin.\cite{mizutani1997direct}
Lian {\em et al.} suggested that fast energy transfer is facilitated by low-frequency collective motions of the protein, which are heavily damped by the surrounding water, enabling efficient energy dissipation.\cite{lian1994energy} 
Lervik {\em et al.} simulated transient non-equilibrium conditions to investigate the heat dissipation from a protein into the surrounding solvent as the system approaches equilibrium, and characterized the thermal conductance of proteins.
Their observations suggest that the protein–water interface plays a major role in the thermal relaxation of biomolecules,\cite{lervik2010heat} supporting the hypothesis that damping of vibrations by interfacial water is crucial for fast solute-to-solvent energy transport.

For simple model systems, {\em e.g.}, 1-D chains of harmonic oscillators, Conti et al. \cite{Conti1d_oscillator} showed that the vibrational density of states (VDoS), when normalized for temperature, exhibits distinct differences between steady-state heat dissipation and equilibrium conditions. 
The differences demonstrate that energy can dissipate at frequency-specific rates between two particular components of a system. 
Along these lines, Niehues {\em et al.} used driven molecular dynamics simulations to show that the efficiency of intermolecular energy transfer from excited intramolecular vibrations in a solute to the solvent depends substantially on frequency.\cite{niehues2012driving} 
Bond vibrations in the mid-infrared dissipate energy into the solvent substantially slower than low-frequency vibrations at terahertz frequencies (far-infrared), whose spectrum overlaps with intermolecular vibrations in the hydrogen bond network of water.

However, several questions remain:
Is the frequency of a vibration and its overlap with the vibrational spectrum of water the key criterion for its ability to transfer energy? 
Are some low-frequency vibrations in proteins more or less efficient in their ability to thermally couple to the surrounding solvent?
Answering such questions is complicated by poor characterizations of low-frequency protein vibrations in proteins, which, especially at far-infrared frequencies (0-200~\wn\ or 0-6~THz), are highly anharmonic in nature ($h \nu < k_B T$).

We recently developed FREquency-SElective ANharmonic (FRESEAN) mode analysis \cite{fresean} to reliably isolate low-frequency vibrations from molecular dynamics simulations. 
In contrast to other methods, FRESEAN mode analysis does not rely on harmonic approximations. 
As a direct consequence, vibrations of a system are not described by a single set of orthonormal modes.
Instead, FRESEAN mode analysis separates collective degrees of freedom (DOF), {\em i.e.}, modes, based on their contribution to the vibrational spectrum (specifically the VDoS) at selected frequencies.\cite{fresean}
Our previous work has shown that, in contrast to harmonic or quasi-harmonic normal mode analysis, FRESEAN mode analysis successfully isolates DOFs associated with low-frequency vibrations,\cite{fresean} which we exploited to enhance the sampling of slow conformational transitions.\cite{mondal2024exploring,sauer2024fast}

Here, we combine FRESEAN mode analysis with all-atom molecular dynamics simulations of a solvated protein to analyze the vibrational energy transfer between a protein and surrounding water as a function of frequency and associated collective DOFs.
For this purpose, we compare simulations under equilibrium and steady-state non-equilibrium conditions.
In the latter case, the protein dissipates energy into the surrounding solvent at a constant net rate.
We demonstrate that this net rate is the sum of two key contributions: a) fast energy transfer mediated by a small number of diffusive modes and damped low-frequency vibrations; and b) vibrational modes that couple to the solvent less efficiently but dominate overall energy transfer by their larger number.

%% file: theory.tex
\section{Theory}
\subsection{Selecting Collective Degrees of Freedom by Frequency}
To analyze kinetic energy distributions among protein vibrational modes as a function of frequency, we first performed FREquency-SElective ANharmonic (FRESEAN) mode analysis~\cite{fresean} for an equilibrium simulation of a protein (here: ubiquitin, see Methods section) in solution.

As a first step, we define weighted velocities for the simulated protein atoms $\tilde{\tx{v}}_{i}(t) = \sqrt{m_i} \tx{v}_{i}(t)$. 
Here, $i$ indicates one of the $3N$ degrees of freedom (DOF) in the protein.
In Cartesian coordinates, the components $x$, $y$, and $z$ of the velocity vector of the $k$'th protein atom correspond to $\tx{v}_{3k}(t)$, $\tx{v}_{3k+1}(t)$, and $\tx{v}_{3k+2}(t)$, and $m_{3k}=m_{3k+1}=m_{3k+2}$ is the atomic mass of atom $k$.
The average kinetic energy of all protein atoms is then simply $E_\tx{kin} = \frac{1}{2} \sum_i^{3N} \left\langle\tilde{\tx{v}}_{i}^2\right\rangle_t$.
Here, $\left\langle...\right\rangle_t$ indicates an ensemble average over the simulation time $t$.

In this study, we avoid the use of bond constraints in our simulations, which would require special treatment to disentangle kinetic energy contributions of fractional atomic DOFs~\cite{sanderson2024local}. 
The only constraints present in our simulations affect the center-of-mass motion of the entire system, which are straightforward to take into account in our analysis (see below).

Following standard FRESEAN mode analysis,\cite{fresean} we then define time correlation functions $C_{\tx{\tilde{v}},ij}(\tau) = \left\langle \tilde{\tx{v}}_{i}(t) \cdot \tilde{\tx{v}}_{j}(t+\tau) \right\rangle_t$ that describe the elements of a correlation time-dependent matrix $\vc{C}(\tau)$, including all auto- ($i=j$) and cross- ($i\ne j$) correlations of weighted velocity components. 
The Fourier transforms of these matrix elements $C_{\tx{\tilde{v}},ij}(\nu)=\int_{-\infty}^{+\infty} \tx{e}^{\tx{i} 2 \pi \nu \tau} C_{\tx{\tilde{v}},ij}(\tau) \,d\tau$ result in a frequency-dependent matrix $\vc{C}(\nu)$ whose trace describes the vibrational density of states (VDoS) of the protein~\cite{fresean}, {\em i.e.}, the frequency-dependent distribution of DOFs and their associated kinetic energy:
\begin{equation}
    I_\tx{VDoS}(\nu) = \frac{2}{k_BT} \sum_i^{3N} C_{\tx{\tilde{v}},ii}(\nu)
    \label{e:vdos}
\end{equation}
The latter is invariant to unitary transformations of $\vc{C}(\nu)$.
Thus, diagonalization of $\vc{C}(\nu)$ at any frequency $\nu$ with $\vc{C}(\nu)=\vc{Q}(\nu) \vc{\Lambda}(\nu) \vc{Q}^T(\nu)$, where the matrix $\vc{Q}(\nu)$ consists of the orthonormal eigenvectors $\vc{q}_i^\nu$ of $\vc{C}(\nu)$ as columns and $\vc{\Lambda}(\nu)$ is the diagonal matrix of the corresponding eigenvalues $\lambda_i(\nu)$, results in an alternative expression for the protein VDoS at frequency $\nu$:
\begin{equation}
    I_\tx{VDoS}(\nu) = \frac{2}{k_BT} \sum_i^{3N} \lambda_i(\nu)
    \label{e:lambda}
\end{equation}
At a selected frequency $\nu = f$, each eigenvalue $\lambda_i(f)$ obtained from $\vc{C}(f)$ thus describes VDoS contributions of fluctuations along the corresponding eigenvector $\vc{q}_i^{f}$ at frequency\,$f$.

Each eigenvector $\vc{q}_i^{f} = \left\{ q_{i,1}^{f}, ..., q_{i,3N}^{f} \right\}$ describes a single collective DOF of the protein ({\em i.e.}, a displacement relative to a reference structure) and the weighted velocity along it is defined by the projection:
\begin{equation}
\dot{q}_i^{f}(t) = \sum_j^{3N} \tilde{\tx{v}}_j(t) \, q_{i,j}^{f}
\label{e:qdot}
\end{equation}
The corresponding kinetic energy is given as $\frac{1}{2}\left(\dot{q}_i^{f}\right)^2$, which, absent any constraints, yields $\frac{1}{2} k_B T$ in equilibrium. 
More generally, we can quantify the effective single DOF temperature for any eigenvector $\vc{q}_i^{f}$ as: $T_{q_i^{f}} = \left\langle \left(\dot{q}_i^{f}\right)^2 \right\rangle_t / k_B$.

However, in MD simulations that remove the center-of-mass motion of the system, eigenvectors that describe rigid body translations of the protein are partially constrained.
In FRESEAN mode analysis of a freely diffusing solute, rigid body translations are described by the first three eigenvectors (largest eigenvalues) of $\vc{C}(f=0)$, {\em i.e.}, at zero frequency.\cite{fresean,sauer2025high}
Due to the center-of-mass constraint applied to the system, these three eigenvectors are partially constrained and contribute only $3 \times \left(1 - \frac{m_\tx{protein}}{m_\tx{total}}\right)$ DOFs to the system.
The effective temperature for each of these partially constrained DOFs thus becomes: $T_\tx{trans}^\tx{protein} =  \left(1 - \frac{m_\tx{protein}}{m_\tx{total}}\right)^{-1} \, \left\langle \left(\dot{q}_{i=[1,2,3]}^{f=0}\right)^2 \right\rangle_t / k_B$, where $i$ describes the index of the zero-frequency eigenvector.

At non-zero frequencies, no specific eigenvector of $\vc{C}(f\neq0)$ isolates rigid-body translations. 
In principle, we can use a simple projection on rigid-body translations to quantify the partial constraint applied to any collective protein DOF described by an eigenvector of $\vc{C}(f\neq0)$. 
However, apart from the first three eigenvectors of $\vc{C}(f=0)$ that isolate rigid-body translations, we find that the resulting corrections are minimal and of no consequence for our analysis of single DOF temperatures.

In addition to the temperature of a given collective DOF described by an eigenvector $\vc{q}_i^{f}$, we can compute its VDoS contributions at all frequencies as the Fourier transform of the corresponding time auto-correlation function:
\begin{eqnarray}
C_{q_i^{f}}(\tau) &=& \left\langle \dot{q}_i^{f}(t) \, \dot{q}_i^{f}(t+\tau)\right\rangle_t 
\label{e:qvacf}\\
I_\tx{VDoS}^{q_i^{f}}(\nu) &=& \frac{2}{k_B T_{q_i^{f}}} \int_{-\infty}^{+\infty} \tx{e}^{\tx{i} 2 \pi \nu \tau} C_{q_i^{f}}(\tau) \, d\tau
\label{e:qvdos}
\end{eqnarray}

However, the information for all frequency-dependent auto- and cross-correlations of weighted atomic velocities is already fully described by the matrix $\vc{C}(\nu)$.
Therefore, we can alternatively rewrite Eq.~\ref{e:qvdos} as a simple vector-matrix-vector product:
\begin{equation}
I_\tx{VDoS}^{q_i^{f}}(\nu) = \frac{2}{k_B T_{q_i^{f}}} \left(\vc{q}_i^{f}\right)^\top \vc{C}(\nu) \, \vc{q}_i^{f}
\label{e:qtCq}
\end{equation}

Notably, as defined above, $I_\tx{VDoS}^{q_i^{f}}(\nu)$ describes the vibrational spectrum for a single unconstrained DOF (partial constraints are considered in our definition of $T_{q_i^{f}}$). 
Thus, integration over all frequencies yields by definition: $\int_0^\infty I_\tx{VDoS}^{q_i^{f}}(\nu) \, d\nu = 1$.

\subsection{Selecting Collective Degrees of Freedom by Temperature}

We can formulate an alternative eigenvalue problem that separates DOFs solely based on their average kinetic energy or temperature alone, independent from frequency-dependent contributions to the VDoS.
This is given by the eigenvalues and eigenvectors of the matrix $\vc{C}(\tau = 0)$, which describes instantaneous or static (mass-weighted) velocity correlations. 
Its eigenvectors $\vc{q}_i^\tx{kin}$ describe collective DOFs with an average temperature $T_i = \lambda_i^\tx{kin}/k_B$ defined by the eigenvalue $\lambda_i^\tx{kin}$.
In an equilibrium system, one would expect these eigenvalues to be identical to the average temperature of the system $T$.
As a consequence, any linear combination of Cartesian DOFs results in an eigenvector of $\vc{C}(\tau = 0)$, {\em i.e.}, the eigenvalue problem is fully degenerate. 

In finite-time simulations of a system with a large number of DOFs, this approach will still distinguish collective DOFs with distinct temperatures due to non-ergodic properties that only vanish for infinite simulation times.
However, the eigenvectors $\vc{q}_i^\tx{kin}$ should still be random linear combinations of Cartesian DOFs with no specific properties. 
The latter can be characterized by the spectrum of fluctuations, {\em i.e.}, contributions to the VDoS, along the corresponding eigenvectors, which is most easily obtained in analogy to Eq.~\ref{e:qtCq}.
\begin{equation}
I_\tx{VDoS}^{q_i^\tx{kin}}(\nu) = \frac{2}{k_B T} \left(\vc{q}_i^\tx{kin}\right)^\top \vc{C}(\nu) \, \vc{q}_i^\tx{kin}
\label{e:qtCq2}
\end{equation}
For an equilibrium system, the VDoS of all eigenvectors $\vc{q}_i^\tx{kin}$ should be the same even if the corresponding eigenvalues indicate deviations from equi-partition due to non-ergodicity.

In contrast, in a non-equilibrium system that intrinsically does not obey equi-partition of kinetic energy, diagonalization of $\vc{C}(\tau = 0)$ will separate collective DOFs systematically based on their effective average temperatures.
If the non-uniform distribution of kinetic energies is related to vibrational properties, the VDoS contributions of the eigenvectors $\vc{q}_i^\tx{kin}$ will reflect the vibrational characteristics of collective DOFs with high and low temperatures, respectively.

%% file: methods.tex
\section{Methods}
\subsection{Simulation Details}
To prepare a system for MD simulations, the protein ubiquitin (PDBID: 1UBQ) was placed in a 80~\AA\ × 80~\AA\ × 80~\AA\ cubic box with periodic boundary conditions and solvated with flexible TIP4P/2005 water~\cite{tip4p_flex}.
All simulations were performed using GROMACS 2022.5 \cite{gromacs} and with the AMBER99SB-ILDN force field for the protein~\cite{amber99sb-ildn}. 
Short-range electrostatic and Lennard-Jones interactions were computed using a 10~\AA\ real-space cutoff with long-range dispersion corrections for energy and pressure. 
Long-range electrostatic interactions were computed with the Particle Mesh Ewald (PME) method \cite{DardenParticleMeshEwald} using a 1.2~\AA\ grid spacing and fourth-order interpolation.
All covalent bonds/angles in the protein and water were treated as flexible to simplify the evaluation of kinetic energies for individual DOFs~\cite{sanderson2024local}. 
Unless noted otherwise, the only constraints applied to our simulations were used to remove the center-of-mass translation of the system.
Prior to dynamic simulations with a time step of 0.5~fs in equilibrium and steady-state non-equilibrium ensembles, we performed an energy minimization of the system using a steepest descent algorithm.

\subsection{Equilibrium Ensemble}
The system was equilibrated for 1000~ps in the isothermal-isobaric ensemble, while the protein and solvent were coupled to separate heat baths with identical temperatures of 300~K.
We used the V-rescale \cite{Vrescale_Bussi} thermostat with a 1~ps time constant and a Berendsen barostat \cite{Berendsen} with a 2~ps time constant.
For production simulations, we switched to the Parrinello-Rahman barostat \cite{nose1983constant,parrinello1981polymorphic} with a 5~ps time constant for pressure coupling. 
To sample the equilibrium velocity time auto- and cross-correlation functions defined above, we simulated this system for 100~ns. 
Atomic coordinates and velocities for the entire system were saved every 100~ps.
To sample fast vibrations of the protein (including vibrations of covalent bonds involving hydrogens), coordinates and velocities of the protein atoms were communicated to MDAnalysis~\cite{michaud2011mdanalysis,gowers2019mdanalysis} every 4~fs via a recently developed streaming interface~\cite{imdclient} in the first 20~ns of the simulation and written to a separate trajectory file.
This high time-resolution trajectory was then used to compute velocity time correlation functions and their Fourier transforms (see Theory section).
For the remaining 80~ns of the production simulation, we reduced the time-resolution of the streamed trajectory to 500~fs.
This intermediate time-resolution was used to compute average temperatures of collective DOFs as well as the matrix $\vc{C}(\tau=0)$.

\subsection{Non-Equilibrium Ensemble}
To establish a steady-state non-equilibrium ensemble with a constant temperature gradient between the protein and the surrounding solvent, we set the temperatures of the thermostats coupled to the protein and water to 310~K and 290~K, respectively, in both the equilibration and production simulations.
Under these conditions, the system establishes a steady state with constant heat flux from the protein to the solvent and an average protein temperature of 306.0~K.
The latter depends on the coupling strength (time constant) of its thermostat and the overall rate of the protein-water heat exchange.
Otherwise, we kept the simulation protocol identical to the equilibrium system.

\subsection{Reference Simulations of Water}
As a pure water reference, we simulated a small cubic box of flexible TIP4P/2005 water with 221 water molecules using the same protocol for equilibration and production simulations as for equilibrium simulations of the solvated protein.

\subsection{Analysis}
All simulations are analyzed in independent 1~ns segments to compute statistical error bars for all computed properties.
Prior to analysis, we rotated protein coordinates (relative to protein center-of-mass) and velocities in our trajectories to minimize the root mean squared deviations with protein coordinates of the energy-minimized structure.
We computed all velocity auto- and cross-correlation functions of weighted atomic velocities described in the Theory section with a maximum correlation time of 2~ps and a time resolution of 4~fs using the first 20~ns of the production simulations.
Using the time symmetry of correlation functions (exact for auto-correlations; implied for cross correlations in equilibrium), this allows for a frequency resolution of 0.25~THz or approximately 8~\wn.
For the equilibrium simulation trajectory, we performed FRESEAN mode analysis following the protocol outlined by Sauer and Heyden \cite{fresean} for all degrees of freedom of ubiquitin (1231 atoms including hydrogens), generating weighted velocity correlation matrices $\vc{C}(\nu)$ with a dimension of $3693 \times 3693$ DOF for 500 frequencies $\nu$.
Diagonalization of $\vc{C}(\nu)$ for a given frequency $\nu=f$ yielded eigenvectors $\vc{q}_i^f$ that describe collective DOFs of the protein, whose contributions to the (equilibrium) VDoS at frequency $f$ are directly proportional to the corresponding eigenvalue (Eq.~\ref{e:lambda}).
Using Eq.~\ref{e:qtCq}, we further used $\vc{C}(\nu)$ to compute the frequency-dependent contributions to the (equilibrium) VDoS of collective DOFs described by selected eigenvectors of $\vc{C}(\nu)$ at multiple frequencies.

Next, we projected weighted atomic velocities in both full-length (100~ns) equilibrium and steady-state non-equilibrium simulation trajectories with 500~fs time resolution on eigenvectors of $\vc{C}(\nu)$ according to Eq.~\ref{e:qdot}. 
These projections allowed us to compute and compare average temperatures in both simulations for the same set of collective DOFs described by the eigenvectors of $\vc{C}(\nu)$.
Further, we used the 100~ns simulations for both the equilibrium and steady-state systems to compute the matrix $\vc{C}(\tau=0)$, which does not require sampling of fluctuations with high time resolution. 
As described in the Theory section, the eigenvalues and eigenvectors of $\vc{C}(\tau=0)$ separate collective degrees of freedom by their average temperature in the simulation.
For the collective DOFs described by both sets of eigenvectors (equilibrium and steady-state), we again computed the frequency-dependent contributions to the equilibrium VDoS using Eq.~\ref{e:qtCq}.

%% file: results_and_disc.tex
\section{Results and Discussion}
\subsection{Vibrational Density of States}

\begin{figure}[ht!]
    \centering
    \includegraphics[width=0.45\textwidth]{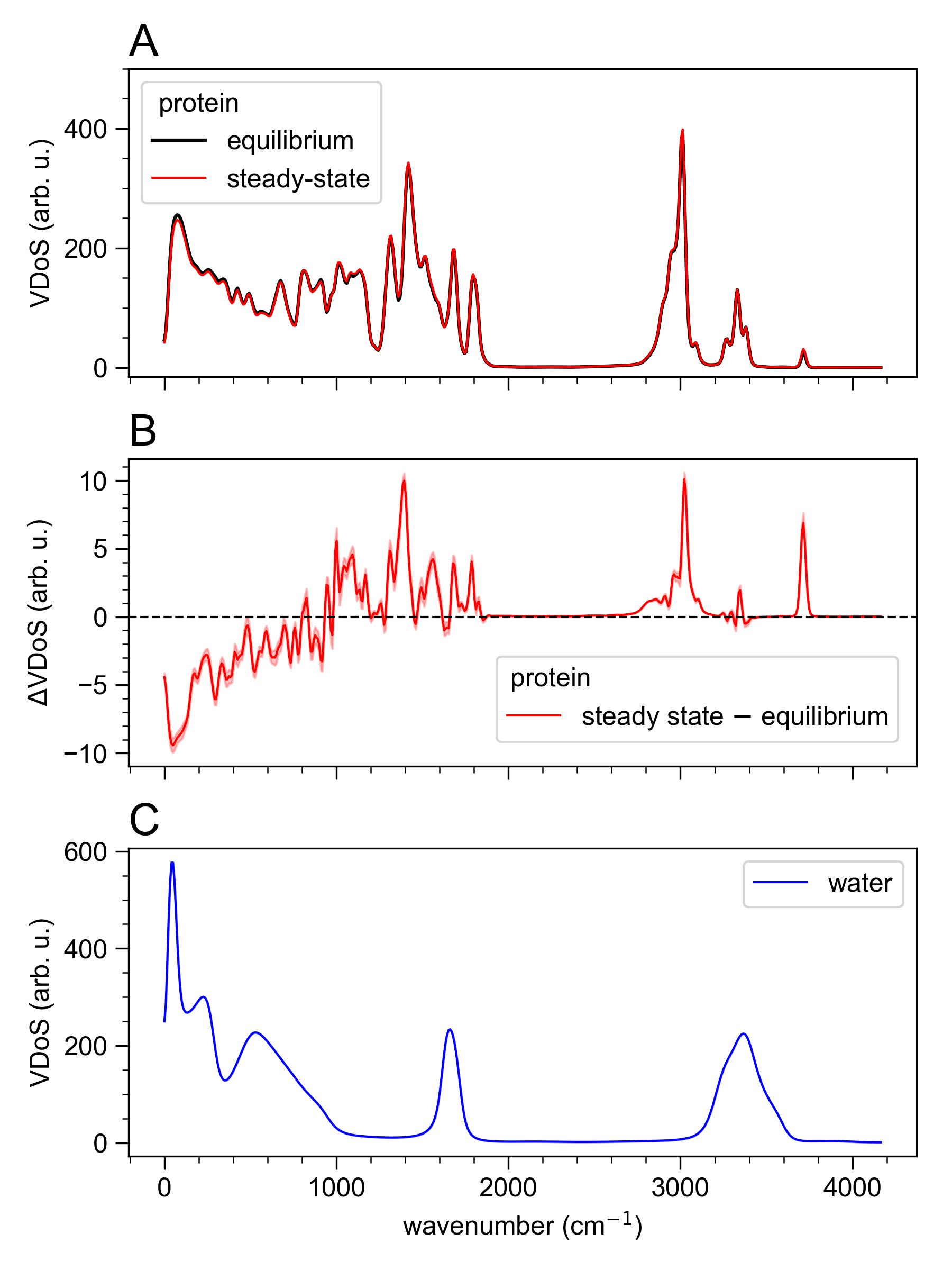}
    \caption{Vibrational density of states (VDoS) of ubiquitin. (A) Comparison of equilibrium (black) and steady-state non-equilibrium (red) simulations with constant heat flux from the protein to the solvent. (B) Difference between the steady-state and equilibrium VDoS from (A). (C) VDoS of (pure) water simulated with the flexible flexible TIP4P/2005 model used as a solvent in the protein simulations. Error bars are shown as shaded regions but are only faintly visible in the difference spectrum in panel B.}
    \label{f:vdos}
\end{figure}

To compare simulations of solvated ubiquitin in equilibrium and steady-state non-equilibrium conditions, we first analyzed the VDoS defined in Eq. \ref{e:vdos}, which we computed directly from mass-weighted time auto-correlations of atomic velocities of the protein in both simulations.
The definition of the VDoS includes a normalization by the average protein temperature (Eq.~\ref{e:vdos}), which is 300.0~K and 306.0~K for the equilibrium and steady-state simulations, respectively. 
This allows us to directly compare frequency-dependent distributions of the kinetic energy described by the VDoS in both systems, as shown in Figs.~\ref{f:vdos}A and B.

The comparison shows that the VDoS of the steady-state system is decreased at frequencies between 0-1000~\wn\ relative to its equilibrium counterpart.
This decrease is most clearly visible in the difference spectrum in Figure~\ref{f:vdos}B.
In this frequency range, the far-infrared or THz spectrum ($\approx$~0-30~THz)~\cite{markelz2022perspective}, the vibrational resonances in the protein VDoS are broad and result in a continuous spectrum.

This result clearly shows that protein vibrations in the far-infrared spectrum exhibit below-average temperatures in the steady-state system. 
This demonstrates that low-frequency vibrations exhibit stronger thermal coupling to the cool solvent, {\em i.e.}, these vibrations transfer energy efficiently to surrounding water molecules.
The decrease is most pronounced at approximately 70~\wn\ near prominent peaks in the protein (Figure~\ref{f:vdos}A) and water VDoS (Figure~\ref{f:vdos}C). 
This suggests, as proposed previously,\cite{niehues2012driving,heyden2013spatial,heyden2014resolving} that matching frequencies, even for anharmonic vibrations at low frequencies, play a critical role for efficient energy transfer.

The decreased low-frequency VDoS intensities in the steady-state system are compensated by an increase at high-frequencies, which indicate protein vibrations with higher-than-average temperatures (note: the integral of the VDoS over all frequencies is, by definition, identical for the equilibrium and steady-state simulations).
Thus, high-frequency vibrations are less strongly coupled to the solvent and kinetic energy is transferred to the solvent more slowly.

The increase in the VDoS intensity, {\em i.e.}, excess of kinetic energy, is particularly pronounced for the highest-frequency peak in the protein VDoS at 3700~\wn, especially in relation to its small intensity in Figure~\ref{f:vdos}A.
Like other vibrations at 3000~\wn\ and beyond, this peak is caused by bond vibrations involving hydrogen atoms. 
Specifically, the 3700~\wn peak is caused by vibrations of covalent O-H bonds in amino acids such as serine, threonine and tyrosine, which can be traced back to the corresponding harmonic force constant of these bonds in the AMBER99SB-ILDN force field~\cite{amber99sb-ildn}.
The pronounced excess of kinetic energy in these protein O-H stretch vibrations can be explained by a comparison to the VDoS of flexible water in Fig.~\ref{f:vdos}C. 
The frequency of protein O-H oscillators exceeds the highest frequency vibrations in the surrounding water and thus impedes vibrational energy transfer.

We note that high-frequency bond vibrations are typically absent in force-field based molecular dynamics simulations due to the use of bond constraints. 
From a quantum-mechanical perspective, the energy required to excite oscillators in the mid-infrared exceeds the thermal energy available by molecular collisions, which justifies the use of constraints to eliminate these degrees of freedom in classical simulations. 
In our simulations, we only apply constraints to remove the center-of-mass motion (translation) of the entire system. 
This affects the center-of-mass motion of the protein by reducing the kinetic energy in the corresponding collective degrees of freedom by a factor $m_p/m_t$, where $m_p$ is the mass of the protein and $m_t$ is the mass of the entire system (see Theory section).
Similarly, bond constraints would reduce the kinetic energy in collective degrees of freedom, albeit in a much more complex manner~\cite{sanderson2024local}. 
Thus, we omitted the use of bond constraints in our protein simulations to simplify our analysis of the partition of kinetic energy among collective degrees of freedom.

For consistency, we paired the fully flexible model of the protein with a flexible model for the surrounding water to facilitate energy transfer between high-frequency protein vibrations with the solvent, which would otherwise be limited by a more extensive mismatch of vibrational frequencies~\cite{niehues2012driving}.

We note in passing that bond force constants in empirical protein force fields such as AMBER99SB-ildn~\cite{amber99sb-ildn} have not been optimized to reproduce actual vibrational spectra of proteins. 
However, this does not impact the main goal of this study, which is to characterize the protein DOFs primarily responsible for protein-water energy transfer.

\subsection{Collective Degrees of Freedom Separated by Frequency}

\begin{figure}[ht!]
    \centering
    \includegraphics[width=0.45\textwidth]{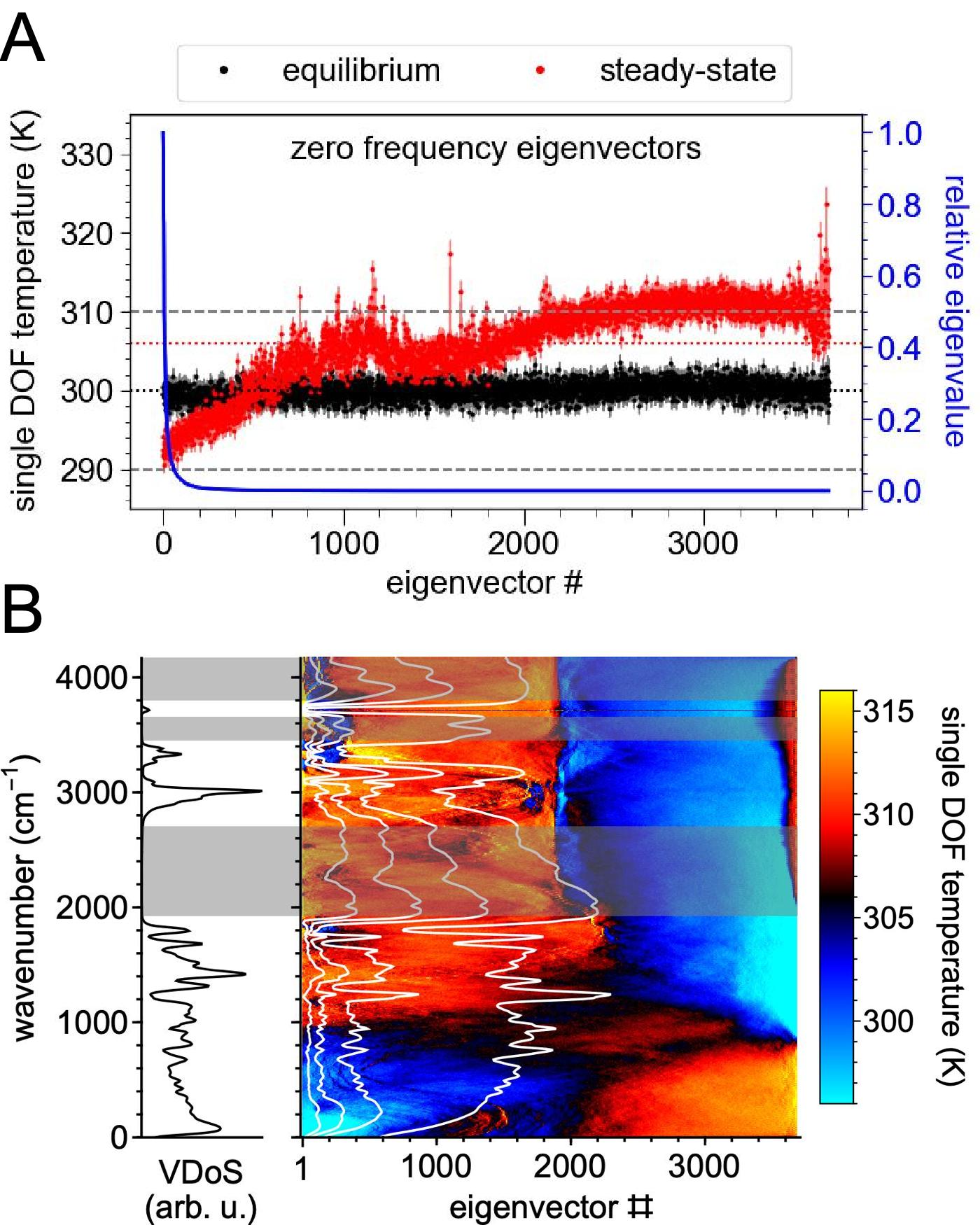}
    \caption{
    Average temperature of collective DOFs obtained from FRESEAN mode analysis of the equilibrium system. 
    (A) Temperatures obtained after projecting the equilibrium (black) and steady-state (red) simulations on all eigenvectors of $\vc{C}(\nu)$ at zero frequency with statistical error bars.
    Dotted horizontal lines in black and red indicate average equilibrium and steady-state protein temperatures.
    Temperatures of thermostats coupled to the protein (310~K) and water (290~K) in the steady-state system are indicated by dashed lines in gray.
    Eigenvectors are sorted by descending eigenvalues (blue) indicated on the alternative y-axis.
    (B) Temperature obtained after projecting the steady-state simulation on all eigenvectors of $\vc{C}(\nu)$ at all sampled frequencies. 
    The color code describes temperatures within $\pm$10~K of the average protein temperature (306~K); superimposed white lines indicate the number of eigenvalues needed to describe 50\%, 75\%, 90\%, and 99\% of the VDoS in Eq.~\ref{e:lambda} for each frequency. 
    The total VDoS as a function of frequency is shown as an inset on the left; shaded in gray are frequencies with insignificant total VDoS intensities.
    }
    \label{f:temp}
\end{figure}

To identify specific low-frequency vibrations of the protein that exhibit below-average temperatures in the steady-state system, we performed FRESEAN mode analysis for the equilibrium system.
The eigenvectors of the frequency-dependent matrix $\vc{C}(\nu)$ for a given frequency $f=\nu$ describe collective DOFs whose contributions to the protein VDoS at frequency $f$ are proportional to their eigenvalues~ (Eq.~\ref{e:lambda}).
By projecting weighted atomic velocities from equilibrium and steady-state simulation trajectories (after rotation into reference orientation; see Methods) on a specific eigenvector using the scalar product (Eq.~\ref{e:qdot}), we analyzed their respective kinetic energies and temperatures.
The results are shown in Figure~\ref{f:temp} for all 3693 eigenvectors obtained at zero frequency (A) as well as for all sampled frequencies (B).

As expected, the temperatures for all vibrational modes in the equilibrium system fluctuate around the corresponding equilibrium temperature at 300~K (shown in black in Figure~\ref{f:temp}A).
However, the same vibrational modes feature distinct temperatures in the steady-state system (red).

To interpret these observations, we note that the VDoS at zero frequency is dominated by contributions from a relatively small number of eigenvectors with non-zero eigenvalues (Eq.~\ref{e:lambda}; relative eigenvalues shown in blue on alternative y-axis of Figure~\ref{f:temp}A). 
In the steady-state system, the temperature of these collective DOFs is very close to the temperature of the surrounding solvent (290~K), indicating strong thermal coupling to the latter.
These modes include the first 6 eigenvectors of $\vc{C}(\nu)$ at zero frequency that describe rigid-body translations and rotations of the protein~\cite{fresean} as well as low-frequency vibrations with resonances on the order of 10-20~\wn\ (see \1{Figure~S1 in the Supplementary Material, SM,} for VDoS and time-dependent velocity correlations along selected DOFs).

The temperature of zero-frequency eigenvectors in the steady-state system increases steadily with their index number and reaches the average protein temperature of 306~K approximately for eigenvector 1000. 
At this point, the eigenvalues are indistinguishable from zero resulting in eigenvectors that are increasingly random linear combinations of the remaining DOF that have only one shared characteristic: they do not contribute to the zero-frequency VDoS.
Notably, for eigenvectors with indices 2000 and above, the average temperatures are close to or even exceed the reference temperature of the protein thermostat 310~K (protein thermostat).
This indicates very weak coupling to the surrounding solvent for collective DOFs that do not contribute to the zero-frequency VDoS.
In the steady-state system, kinetic energy is provided by the protein thermostat uniformly to all protein DOFs and builds up in DOFs that cannot efficiently transfer this energy into the solvent.
The heat dissipation rate increases with the temperature differential $\Delta T$ to other DOFs in thermal contact until kinetic energy inputs and outputs are equal for each DOF (steady-state condition). 

In Figure~\ref{f:temp}B, we extend our observations to eigenvectors of $\vc{C}(\nu)$ obtained at all sampled frequencies.
At low frequencies (up to approximately 1000~\wn), eigenvectors with significant VDoS contributions at that frequency (low index, large eigenvalue) are cold with temperatures close to the solvent.
As observed in Figure~\ref{f:temp}A, the coldest collective DOF are the ones with the largest contributions to the zero-frequency VDoS (smallest index).
Eigenvectors at frequencies below 1000~\wn\ with indices of 2000 and higher (negligible eigenvalues) effectively do not contribute to the low-frequency VDoS and instead feature high temperatures close to or exceeding the protein thermostat reference temperature. 

This behavior reverses for vibrations at frequencies exceeding 1000~wn. 
Collective DOF contributing to the VDoS at high frequencies (low index, large eigenvalue) are hot, while eigenvectors with zero eigenvalues (large index) tend to be cold (exceptions are the highest-indexed eigenvectors that describe orthogonal harmonic at other frequencies in the high-frequency regime).

We note that frequencies with no significant protein VDoS intensities (shaded in gray in Figure~\ref{f:temp}B) can be ignored in this analysis, even though the results for those frequencies follow the same pattern.

A detailed analysis of the relationship between eigenvectors of $\vc{C}(\nu)$ at different frequencies (\1{Figure~S2 in the SM}) reveals that the low-temperature eigenvectors obtained at zero frequency have significant similarities with eigenvectors obtained at other low frequencies, in no small part due to the broad lineshape of vibrations along any of the low-frequency vibrations (see \1{Figure~S1A of the SM}). 
The latter implies that eigenvectors with large VDoS contributions at zero frequency have non-zero contributions to the VDoS at, {\em e.g.}, 100~\wn\ and vice versa.

Overall, the results in Figure~\ref{f:temp} demonstrate that collective protein DOFs that contribute to the protein VDoS at low frequencies (including rigid-body diffusion) feature temperatures that are essentially identical to the cold solvent in the steady-state system.

\subsection{Collective Degrees of Freedom Separated by Temperature}

\begin{figure}[ht!]
    \centering
    \includegraphics[width=0.45\textwidth]{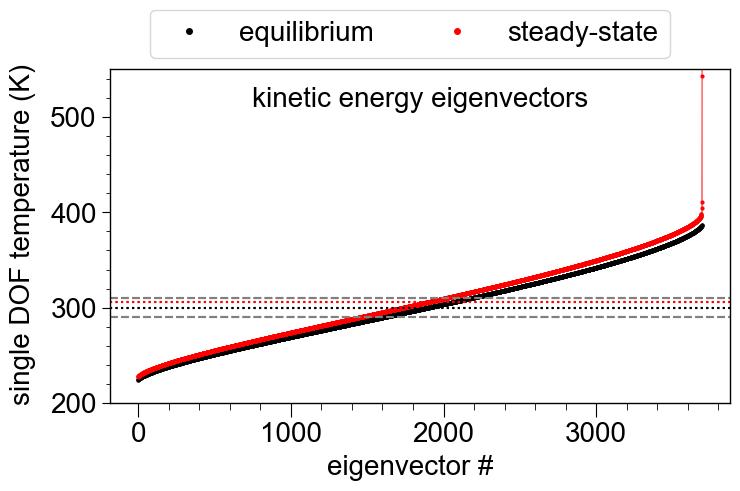}
    \caption{
    Average temperature of collective DOFs constructed specifically based on differences in kinetic energy as eigenvectors of the static velocity correlation matrix $\vc{C}(\tau = 0)$. 
    Results for the equilibrium and steady-state simulations are shown as black and red dots, respectively.
    Statistical errors are shown as shaded areas but negligible on the scale shown.
    Average protein temperatures in the equilibrium and steady-state simulations are indicated as dotted horizontal lines in the respective colors.
    The temperatures of the thermostats coupled to the protein (310~K) and water (290~K) in the steady-state simulation are indicated by horizontal dashed lines in gray.
    }
    \label{f:kin}
\end{figure}

In Figure~\ref{f:kin}, we also analyze the temperature of individual collective DOF.
However, here we select these collective DOFs specifically for their temperature by diagonalizing the static velocity correlation matrix $\vc{C}(\tau=0)$, which we computed for both the equilibrium and steady-state simulations.
The average temperatures of projected simulation trajectories are directly proportional to the eigenvalues of $\vc{C}(\tau=0)$ (sorted here in ascending order).
Compared to Figure~\ref{f:temp}, it is striking that the temperature variations are substantially larger in Figure~\ref{f:kin}.
The reason is that we constructed the collective DOFs here specifically to separate DOFs with distinct temperatures. 
Even in the equilibrium system, we observe collective DOFs with temperatures as low as 224~K and as high as 387~K.
For the steady-state system, we observe a not-quite uniform shift to higher temperatures for most eigenvectors and a small number of hot DOFs with temperatures up to 540~K.

The first question that arises in face of the result in Figue~\ref{f:kin} is whether the variation of single DOF temperatures, in particular in the equilibrium system where equi-partition of kinetic energy is expected, can be explained by statistical noise and finite-time sampling, {\em i.e.}, non-ergodic behavior. 
To investigate this, we performed the same analysis as a function of the trajectory length in \1{Figure~S3 of the SM} and observed a clear dependence of the width of the distribution with sampling time.
With increasing sampling time, the width of the distribution of single DOF temperatures decreases, consistent with a decrease in statistical noise.
This analysis reveals that it takes a substantial amount of simulation time to achieve ergodic behavior and energy equi-partition in simulations of high-dimensional systems such as proteins. 
In fact, the analysis of eigenvalues of the matrix $\vc{C}(\tau = 0)$ introduced here may be used as a highly sensitive measure to quantify deviations from ergodicity in future applications. 

A potential remedy of the slow equi-partition of kinetic energy detected here, apart from substantially longer simulation times, may be the use of so-called massive thermostats during equilibration that are coupled to each individual DOF and are used frequently in {\em ab initio} molecular dynamics simulations~\cite{marx2009ab}.
In the current case, our results in Figure~\ref{f:kin} and \1{Figures~S3 of the SM} indicate that simulations on the microsecond timescale would be required to reduce the spread of single collective DOF temperatures to less than 10~K using a standard global thermostat.
While technically feasible, our further analysis below does not indicate that an extension of the trajectories would change the outcome of our analysis.

\begin{figure}[ht!]
    \centering
    \includegraphics[width=0.45\textwidth]{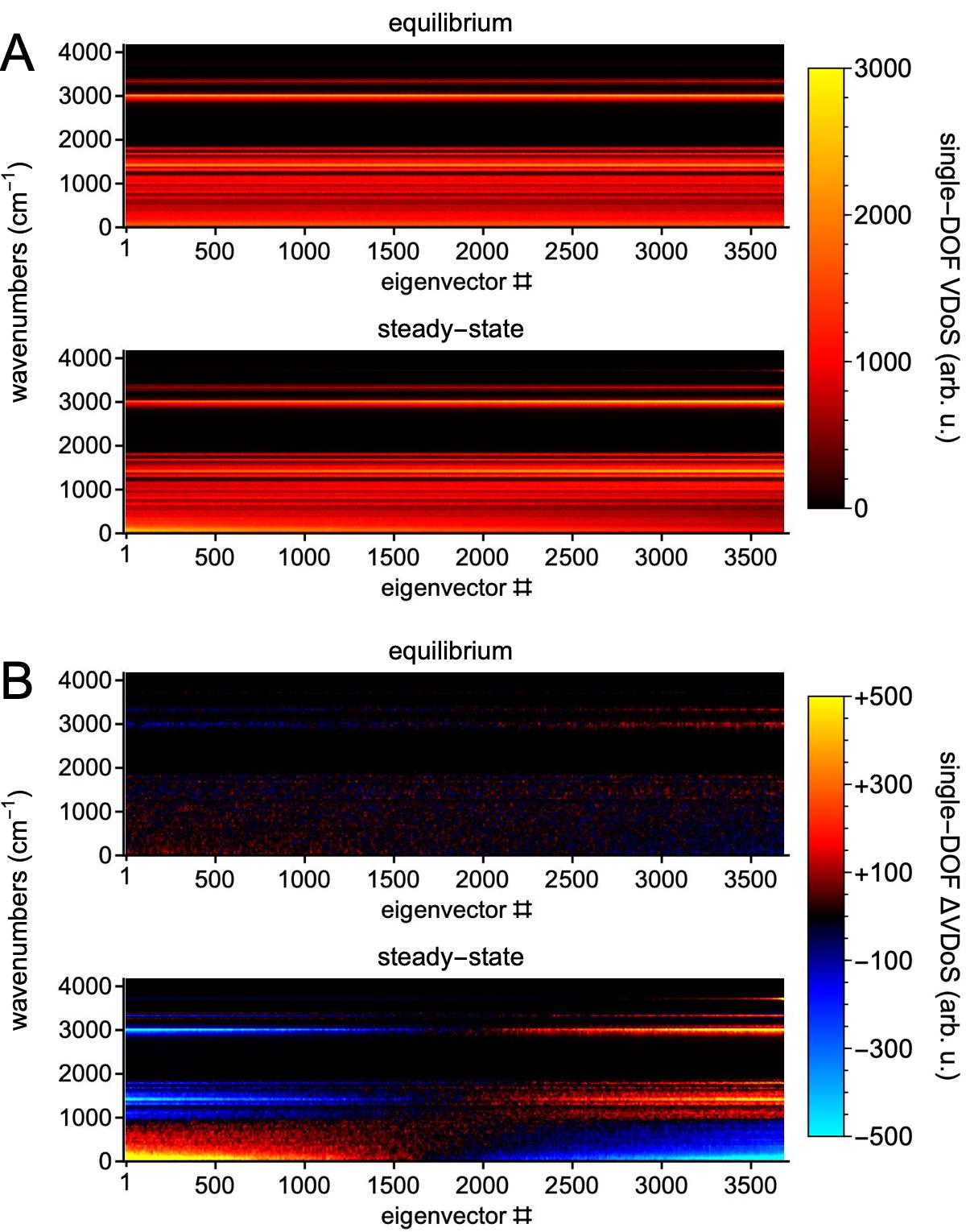}
    \caption{VDoS of single collective DOFs. 
    (A) VDoS of fluctuations along eigenvectors of the static velocity correlation matrix $\vc{C}(\tau = 0)$ obtained from the equilibrium (top panel) and steady-state (lower panel) simulations.
    (B) Difference of the single DOF VDoS with respect to the average VDoS over all DOF (the average is equivalent to the protein VDoS for the equilibrium system shown in Figure~\ref{f:vdos}).}
    \label{f:qvdos}
\end{figure}

Specifically, we ask the second question whether the low- and high-temperature DOFs identified in Figure~\ref{f:kin} have distinguishable properties. 
For this purpose, we computed the frequency-dependent VDoS contributions of each of the eigenvectors of $\vc{C}(\tau = 0)$ (computed from equilibrium and steady-state simulations) using Eq.~\ref{e:qtCq2}.
This is equivalent to projecting the equilibrium simulation trajectory onto selected collective DOFs and analyzing the resulting fluctuations as described in Eq.~\ref{e:qvdos}.
Therefore, the analyzed trajectory is identical and differences between VDoS contributions of collective DOFs are fully attributable to the latter, {\em i.e.}, differences between the eigenvectors of $\vc{C}(\tau = 0)$ obtained from the equilibrium and steady-state simulations.
The results are shown in Figure~\ref{f:qvdos}.

In Figure~\ref{f:qvdos}A, we recognize the peak intensities of the total VDoS in Figure~\ref{f:vdos}.
For eigenvectors obtained from the equilibrium system shown in the top panel, the VDoS contributions of the individual DOFs are essentially identical and no trend with eigenvector index (or temperature as shown in Figure~\ref{f:kin}) is apparent.
This is confirmed in the more sensitive representation in the top panel of Figure~\ref{f:qvdos}B, where the difference relative to the average equilibrium VDoS is shown. 
Thus, apart from the differences in average temperature, there is nothing special about the eigenvectors of the static velocity correlation matrix $\vc{C}(\tau =0)$ obtained for the equilibrium system.
This confirms that the temperature variations in Figure~\ref{f:kin} are statistical in nature and simply require longer sampling times to reach equi-partition of kinetic energy in the ergodic limit.

This is quite different for the eigenvectors obtained from the steady-state system.
Careful investigation of the VDoS in the lower panel of Figure~\ref{f:qvdos}A indicates a tendency of low-frequency VDoS intensities (below 500~\wn) to decrease with increasing mode index, while the VDoS of high frequency vibrations follows the opposite trend.
The difference spectrum relative to the average equilibrium VDoS in the lower panel of Figure~\ref{f:qvdos}B highlights this trend in more detail.
Low-indexed eigenvectors associated with lower temperatures clearly feature significantly increased VDoS intensities at low frequencies and decreased intensities at high frequencies.
This is mirrored by an inverse trend for high-indexed eigenvectors associated with high temperatures. 

This analysis also reveals which vibrations are associated with the extreme temperatures of $>$400~K in the steady-state system seen in Figure~\ref{f:kin}.
Isolating the VDoS for the last eigenvectors of the steady-state system reveals spectra dominated by a single peak at 3700~\wn\ associated with the O-H stretch vibration of serine, threonine and tyrosine side chains (\1{Figure~S4 in the SM}).
Thus, while all other protein DOFs, in particular low-frequency vibrations at frequencies below 500~\wn, are able to dissipate energy into the solvent, these O-H vibrations create an internal heat sink. 
Not only are these vibrations unable to effectively transfer energy into the solvent, but their high frequency compared to other protein vibrations also prevents them from transferring energy to other protein DOFs.
Consequently, the temperature of the O-H stretch vibrations increases until $\Delta T$ compensates for inefficient energy transfer mechanisms to other DOFs.

\subsection{Frequency-Dependent Heat Transfer}

Figure~\ref{f:temp} provides clear evidence that collective protein DOFs that are coupled most strongly to the solvent also contribute most to the zero-frequency VDoS. 
This includes rigid-body protein translations and rotations as well as damped low-frequency vibrations.
Specifically, the first eigenvectors of $\vc{C}\left(\nu\right)$ at zero frequency exhibit temperatures close to the solvent thermostat in the steady-state system, despite being thermally coupled to the protein thermostat at a higher temperature. 
Upon further consideration, the connection between the zero-frequency VDoS and thermal coupling to the solvent is intuitive. 
The zero-frequency regime describes non-oscillatory dynamics such as diffusion and the damping of vibrations, both of which are the result of friction. 
In its essence, friction describes loss of kinetic energy to the environment. 
Thus, finding that collective DOFs associated with the zero-frequency VDoS are most strongly coupled to the solvent is a logical consequence.

Protein diffusion requires solvent displacement and disruption of protein-solvent and solvent-solvent interactions, which is the origin of the friction term in the equation of motion for Brownian dynamics.\cite{ermak1978brownian} 
This applies for both translational and rotational diffusion (unless proteins are approximated as non-interacting perfect spheres). 
Likewise, it is reasonable to assume that vibrations that experience friction ({\em i.e.}, damping), and thus contribute to the zero-frequency VDoS, similarly involve solvent displacement and disrupted interactions.
This is especially expected for vibrations that change the conformation of a protein and therefore its shape and solvent interactions. 
This interpretation is supported by the successful use of low-frequency vibrations, selected for their contribution to the zero-frequency VDoS, as collective variables for enhanced sampling simulations of conformational transitions in our previous work.\cite{mondal2024exploring,sauer2024fast}

However, despite our observation that the thermal coupling to the solvent is strongest for collective DOFs associated with the zero-frequency VDoS, we find that the difference between the steady-state and equilibrium VDoS of the protein in Figure~\ref{f:vdos}B exhibits its main negative peak at non-zero frequencies.
The latter roughly corresponds to the lowest-frequency peaks in the protein and water VDoS, which in water is associated with collective hydrogen bond bending vibrations~\cite{walrafen1990raman,heyden2010dissecting}.
In previous work by others and ourselves, efficient solute-solvent energy transfer and correlated vibrational motion have been associated with overlapping vibrational frequencies.\cite{niehues2012driving,heyden2013spatial,heyden2014resolving}
This notion is confirmed by the inability of protein O-H stretch vibrations at 3700~\wn\ to dissipate their vibrational energy.
Nevertheless, our observations in Figure~\ref{f:temp} also show that this picture is incomplete and does not consider energy transfer associated via friction for diffusive motion and dampened low-frequency vibrations.

A third obvious parameter is the number of collective DOFs in the protein associated with a specific energy transfer mechanism. 
To resolve such distinct contributions to protein-water energy transfer, we define here the parameter $\Theta(\nu)$, which describes how much the temperature of protein vibrations at a given frequency differs from $T_\tx{prot}^\tx{ref}$, the reference temperature of the protein thermostat in the steady-state simulation ({\em i.e.}, the expected temperature for all protein DOFs in the absence of heat transfer to the solvent). 
\begin{equation}
\Theta(\nu) = \frac{1}{2}\sum_{i}^{3N} \lambda_{i}(\nu) \cdot \frac{T_\tx{prot}^\tx{ref} - \langle T_{q_i} \rangle}{T_\tx{prot}^\tx{ref}}
\label{e:flux}
\end{equation} 
In steady-state conditions, the protein thermostat adds kinetic energy to all protein DOFs at a constant average rate, which depends on the average steady-state protein temperature (specifically its difference from the thermostat reference).
The closer the temperature $\langle T_{q_i} \rangle$ for a given mode (eigenvector of $\vc{q}_{i}$ of $\vc{C}(\nu)$) is to the solvent temperature, the more efficient is its heat dissipation mechanism.
If $\langle T_{q_i} \rangle$ is larger than the average steady-state protein temperature, $\langle T_\tx{prot} \rangle$ = 306.0~K, its heat dissipation mechanism is less efficient than the average over all protein DOFs.
In the most extreme cases, $\langle T_{q_i} \rangle$ is larger than $T_\tx{prot}^\tx{ref}$ and thus results in negative contributions to $\Theta(\nu)$.

The eigenvalue $\lambda_{i}(\nu)$ describes the contributions of a specific mode to the VDoS at frequency $\nu$ and is thus used as a weighting factor in Eq.~\ref{e:flux} during the summation over all modes at a given frequency. 
As a result, $\Theta(\nu)$ provides a frequency-resolved measure of the efficiency of protein-solvent energy transfer mechanisms. 
By dividing $\Theta(\nu)$ by the total VDoS at a given frequency, we can further obtain the same measurement per degree of freedom. 

\begin{figure}[ht!]
\centering
{\includegraphics[width=0.45\textwidth,scale=1.3]{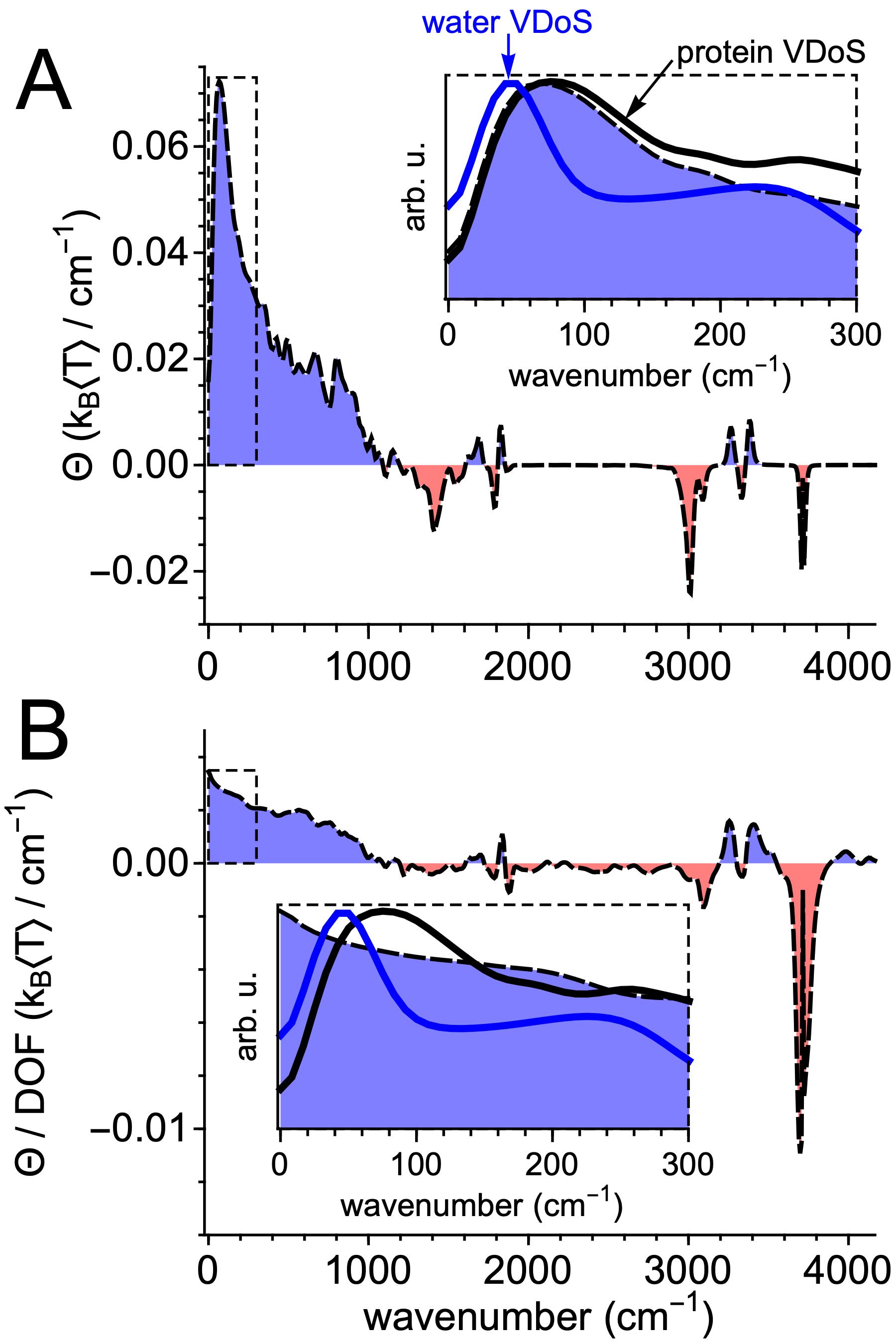}}
\caption{
Protein-solvent heat flux under steady-state conditions. (A) Frequency-resolved heat flux as described by $\Theta(\nu)$ defined in Eq.~\ref{e:flux}. (B) Frequency-resolved heat flux per degree of freedom obtained by dividing $\Theta(\nu)$ by the protein VDoS. Insets: Frequency-resolved heat flux (defined as in corresponding main panel) at low frequencies (dashed boxes) with comparison to scaled protein and water VDoS (scaled to match peak intensity).
}
\label{f:flux}
\end{figure}

The results are shown in Figure~\ref{f:flux}. 
In panel A, the frequency-dependent energy transfer described by $\Theta(\nu)$ is shown and, in the inset, compared to the total VDoS of the protein and water (both normalized by peak intensity) for low frequencies.
$\Theta(\nu)$ peaks at approximately 70~\wn, very close to the low-frequency peak in the protein VDoS.
Notably, the low-frequency peak in the VDoS of water is red-shifted by about 20~\wn. 
This indicates that overlap of vibrational frequencies is not the key criterion for protein-solvent energy transfer.
In that case, one would expect the maximum of $\Theta(\nu)$ at an intermediate frequency between the peaks of the protein and water VDoS.

In panel B, the same comparison is made for $\Theta(\nu)$ per DOF, which confirms that mode per mode, protein-water energy transfer is largest at zero frequency, as discussed earlier. 
However, the small number of modes with significant contributions to the zero-frequency VDoS limits their role in the total ability of the protein to dissipate energy.
Individual vibrations at  70~\wn\ are less effective at dissipating energy, but their large number generates the maximum in the overall frequency-resolved energy transfer efficiency described by $\Theta(\nu)$ in Figure~\ref{f:flux}A.

\subsection{Summary}

Based on our combined observations, two distinct mechanisms govern the heat transfer from the protein to the surrounding solvent. 
The fastest energy transfer occurs for collective DOFs contributing to the dynamics at zero frequency, specifically rigid-body translations/rotations and damped low-frequency vibrations. 
In our steady-state simulations, the corresponding eigenvectors of $\vc{C}(\nu)$ feature temperatures close to the 290~K of the solvent despite being coupled to a thermostat at 310~K.
The heat-transfer in this case can be most easily understood in terms of friction associated with the viscosity of the solvent.

At non-zero frequencies, the heat transfer rate for individual DOFs, {\em i.e.}, vibrational modes, is less effective. 
Nevertheless, the number of modes, specifically close to the peak of the protein VDoS, results in a maximum protein-solvent energy transfer at non-zero frequencies close to 70~\wn.
While friction will play a minor role for these low-frequency vibrations (which are not fully orthogonal from zero-frequency modes, see \1{Figure~S2 in the SM}), the overlapping protein and water VDoS suggests vibrational energy transfer as the primary mechanism.

For higher frequencies, a turning point is reached at approximately 1000~\wn. 
Here, mode temperatures are approximately equal to the average protein temperature (Figure~\ref{f:temp}B).
Vibrational modes at this frequency dissipate energy into the solvent (per DOF) with the same efficiency as the protein on average.

With increasing frequency, the steady-state temperature of vibrational modes increases above $\langle T_\tx{prot} \rangle$ and even exceeds the $T_\tx{prot}^\tx{ref}$.
The corresponding vibrations can only dissipate the energy provided by the protein thermostat after generating a positive temperature gradient with other protein DOFs and the solvent.

Overlapping vibrational frequencies between the protein and solvent are a required factor for energy transfer. 
However, the difference between zero-frequency and non-zero-frequency modes indicates that this is not the only criterion.
Specific characteristics of a collective DOF as well as the number of DOFs are equally important.
However, in the absence of frequency overlap, as seen for protein O-H stretch vibrations at 3700~\wn, heat transfer is extremely inefficient (negative $\Theta(\nu)$ in Figure~\ref{f:flux}).
We identify additional protein vibrations at 1400 and 3000~\wn\ in Figure~\ref{f:flux}, for which $\Theta(\nu)$ is also negative.
A comparison to Figure~\ref{f:vdos}C shows that the protein vibrations at 1400 and 3000~\wn\ also lack overlap with the water VDoS in our simulation.

%% file: conclusion.tex
\section{Conclusion}
Non-equilibrium properties and persistent conversions of energy are a hallmark of living systems.
The cellular metabolism features a large set of reactions and molecular processes that inter-convert chemical, electrical, and mechanical energy.
Locally, these processes generate heat that is than dissipated within the system.
Thermal heat transport in liquids and within proteins has been studied for some time.\cite{leitner2008energy}
However, the detailed mechanisms that govern the heat transfer from biomolecules such as proteins to their solvation environment is less well understood. 
This results in open questions such as: Are there design rules that can increase the ability of a protein to dissipate energy into the solvent? 
The latter could be relevant as a protective mechanism for biomolecules involved in the conversion of large energies, for example, during photosynthesis.

Here, we combine steady-state non-equilibrium simulations of a simple protein with the recently developed FRESEAN mode analysis of biomolecular vibrations.
The latter provides detailed insights into collective degrees of freedom and their contributions to the vibrational spectrum at any frequency sampled in the simulation.
The vibrational spectrum is analyzed in terms of the  VDoS, which directly reports on the frequency-dependent kinetic energy present in vibrations and thus is inherently related to temperature.
The approach is not limited by harmonic approximations and thus allows us to analyze not only oscillatory vibrations but also diffusive motion and damped vibrations at zero frequencies.

Our analysis of the temperature of vibrational modes at distinct frequencies shows that the most efficient protein-solvent heat transfer occurs for diffusive rigid-body motions and solvent-damped protein vibrations. 
Both types of motions contribute to the zero-frequency VDoS and their associated heat transfer mechanism can be understood in terms of solvent friction.

However, the number collective DOFs associated with rigid-body movements and damped vibrations is limited and total energy transfer is dominated by vibrations at non-zero frequencies in the far-infrared.
Specifically, based on the change in the overall VDoS and our own frequency-resolved measure of protein-solvent energy transfer, the latter is most efficient overall for vibrations at approximately 70~\wn\ for our system.
The less efficient vibrational energy transfer compared to the friction-based mechanism at zero-frequency is easily overcompensated by the large number of protein vibrational modes at this frequency.

Overlap between the frequencies of protein and water vibrations appears to be a required condition for efficient energy transfer.
Energy transfer into solvent is extremely inefficient for protein vibrations at frequencies that have no matching resonances in water. 
In our steady-state simulations, the temperature of these vibrations even exceeds the reference temperature of the protein thermostat although the protein is surrounded by a cold solvent.

A modified eigenvalue problem within the framework of FRESEAN mode analysis further allows us to isolate collective protein DOFs by their respective temperature.
Interestingly, this simple analysis provides us with a highly sensitive measure of ergodicity in equilibrium simulations as it reveals the timescales required for true equi-partition of kinetic energy.
While our 100~ns simulations clearly do not achieve the ergodic limit, the vibrational spectra associated with hot and cold DOFs in the equilibrium simulation are essentially identical.
In non-equilibrium steady-state simulations we find instead that the VDoS of cold protein DOFs (which transfer energy to the solvent very efficiently) are dominated by low-frequency vibrations, while the inverse is true for hot protein vibrations.

Future work will analyze individual vibrational modes of the protein in more detail. 
Specifically, we aim to predict the ability of vibrational modes to transfer energy into the solvent via a friction-based mechanism. 
Revealing the relationship between solvent friction and changes in protein-water interactions for low-frequency vibrations may enable us to optimize the ability of proteins to dissipate heat into their environment, or more importantly, to identify proteins that have been optimized for that purpose by evolution.
Further, of course, it would be interesting to analyze the DOFs in the surrounding hydration water that are most susceptible to accept thermal energy from the protein.

%% file: main.bbl
%

%% file: suppinfo.tex
\pagebreak
\begin{widetext}
\newpage
\begin{center}
\textbf{\large Supplementary Material: Protein-Water Energy Transfer via Anharmonic Low-Frequency Vibrations}
\end{center}
\setcounter{equation}{0}
\setcounter{figure}{0}
\setcounter{table}{0}
\setcounter{page}{1}
\makeatletter
\renewcommand{\theequation}{S\arabic{equation}}
\renewcommand{\thefigure}{S\arabic{figure}}
\renewcommand{\thetable}{S\arabic{table}}
\renewcommand{\thepage}{S\arabic{page}}

\subsection{VDoS and VACF of Selected Eigenvectors of $\vc{C}(\nu)$}

\begin{figure}[ht!]
    \centering
    {\includegraphics[width=0.5\textwidth]{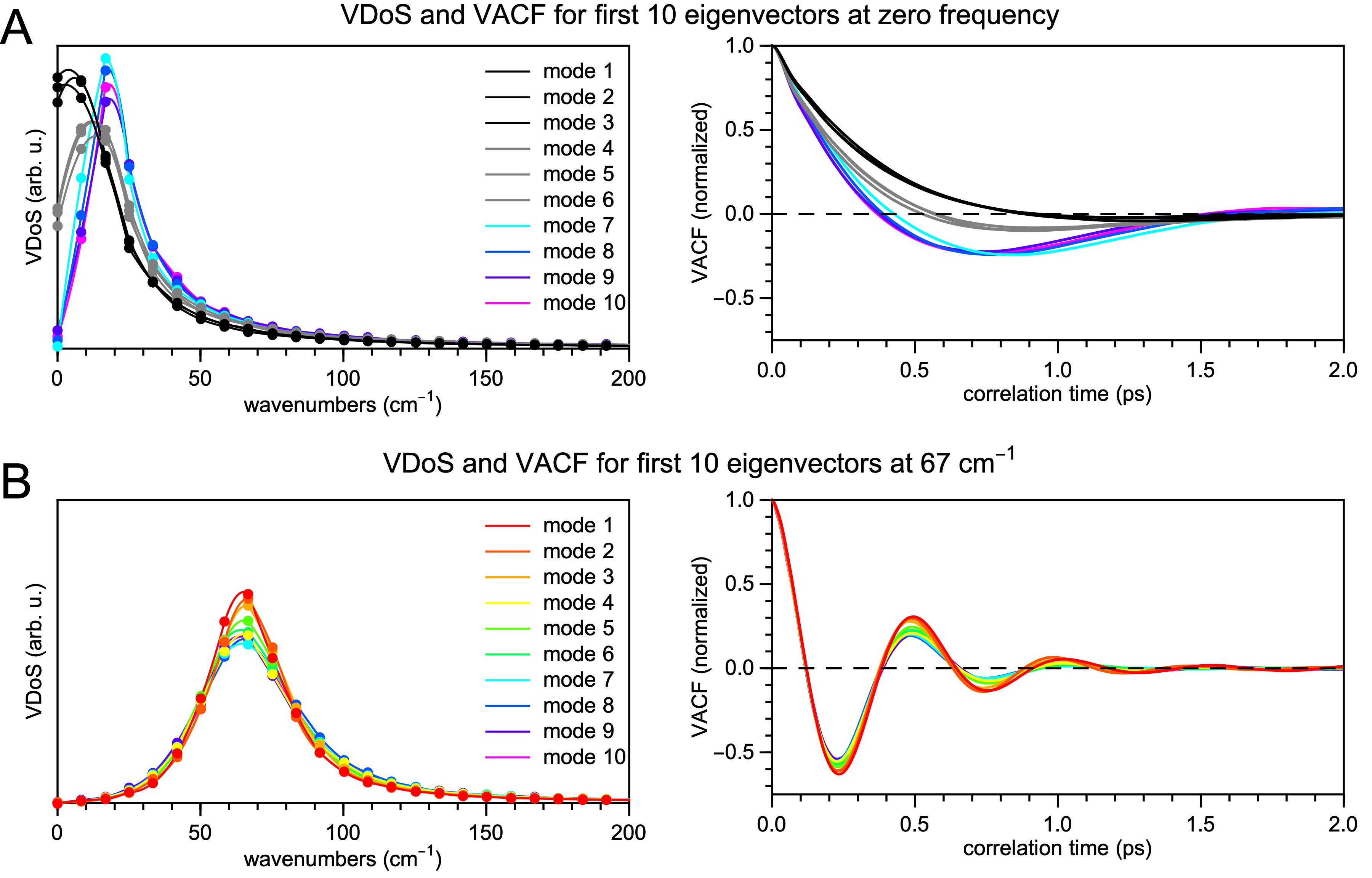}}
    \caption{Fluctuations along selected eigenvectors of $\vc{C}(\nu=f)$ in the equilibrium system. VDoS (left panels) and velocity auto correlation functions (VACF, right panels) for the first 10 eigenvectors of $\vc{C}(f=0)$ (A) and $\vc{C}(f=67 \,\mathrm{cm^{-1}})$ (B). In panel A, we plotted the data for zero-frequency eigenvectors 1-3 (translations) in black and for eigenvectors 4-6 (rotations) in gray. Colors for other eigenvectors are indicated in the legend of the VDoS plots.}
    \label{f:single}
\end{figure}

\subsection{Frequency-Evolution of Vibrational Modes}

A special feature of FRESEAN mode analysis is that it provides a complete set of 3$N$ collective DOFs for every sampled frequency.
The collective DOFs obtained at one frequency are orthogonal and sorted by their contribution to the VDoS at that frequency. 
These DOFs, {\em i.e.}, the eigenvectors of $\vc{C}(\nu)$, evolve smoothly with frequency which corresponds to a rotation in the 3$N$-dimensional space that maintains the separation of large and small VDoS contributions.
Especially at low frequencies, vibrational resonances are not limited to a single frequency but each vibration features a broad continuous spectrum.
This can be observed in our frequency-dependent analysis of VDoS contributions for fluctuations along selected eigenvectors in \1{Figure~\ref{f:single}}.
Consequently, eigenvectors of $\vc{C}(\nu)$ that describe a collective motion whose VDoS contributions exhibit a peak at zero frequency also contribute to the VDoS at non-zero frequencies and vice versa.
This results in a non-random relationship between the eigenvectors of $\vc{C}(\nu)$ for nearby frequencies. 

This is visualized in Figure~\ref{f:corr}, where we calculate the correlation or cosine similarity between the first 1000 eigenvectors of $\vc{C}(\nu)$ at zero frequency with eigenvectors obtained at frequencies corresponding to 50, 100, and 200~\wn.
The correlation between two normalized eigenvectors $\vc{q}_i^{f_1}$ and $\vc{q}_j^{f_2}$ is simply computed as the absolute value of the scalar product.

\newpage

\begin{figure}[ht!]
    \centering
    {\includegraphics[width=0.5\textwidth]{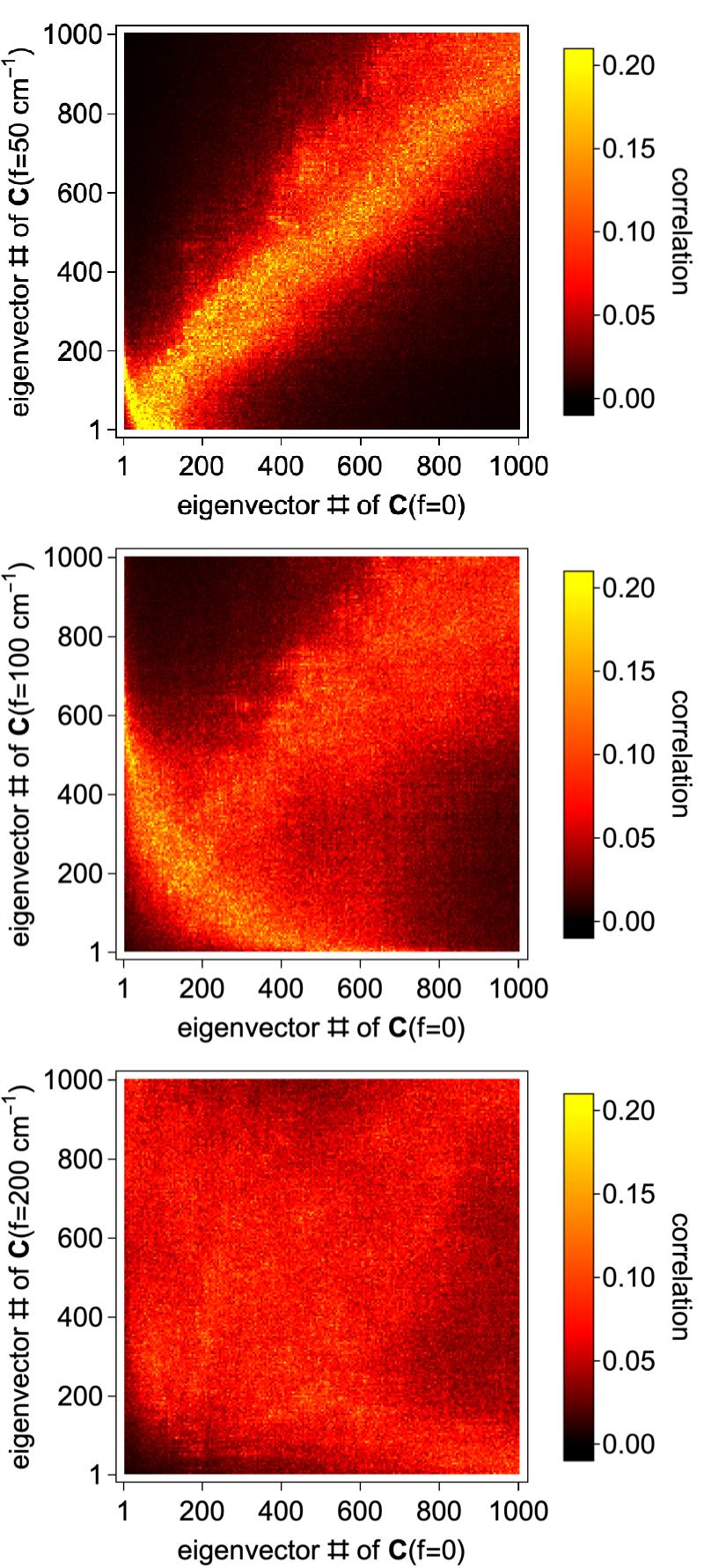}}
    \caption{Correlation between eigenvectors of $\vc{C}(\nu = f)$ at distinct frequencies. Shown are correlations between the first 1000 eigenvectors of $\vc{C}(f = 0)$ and $\vc{C}(f = 50 \, \mathrm{cm^{-1}})$ (A), $\vc{C}(f = 100 \, \mathrm{cm^{-1}})$ (B), and $\vc{C}(f = 200 \, \mathrm{cm^{-1}})$ (C).}
    \label{f:corr}
\end{figure}

\newpage

\subsection{Non-ergodic Behavior as a Function of Simulation Time}

\begin{figure}[ht!]
    \centering
    {\includegraphics[width=0.5\textwidth]{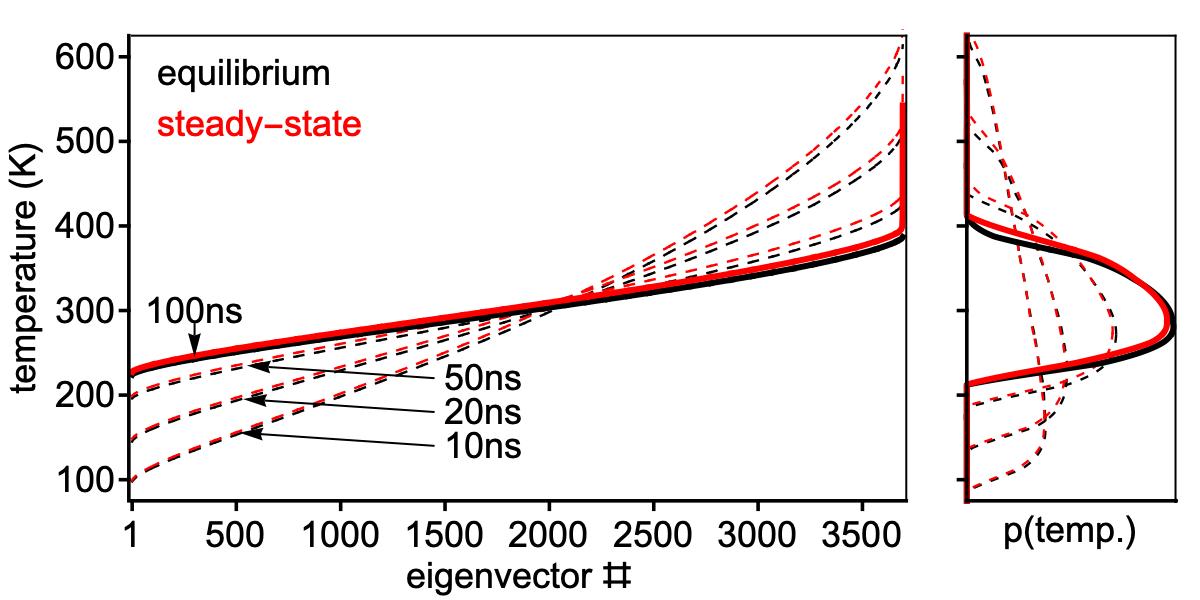}}
    \caption{
    Same analysis as presented in Figure~3 of the main text as a function of the analyzed trajectory length. 
    The inset on the left indicates the corresponding distributions of mode temperatures.
    }
    \label{f:ergodic}
\end{figure}

\subsection{VDoS of Hot Vibrations in the Steady-State Simulation}

\begin{figure}[ht!]
    \centering
    {\includegraphics[width=0.5\textwidth]{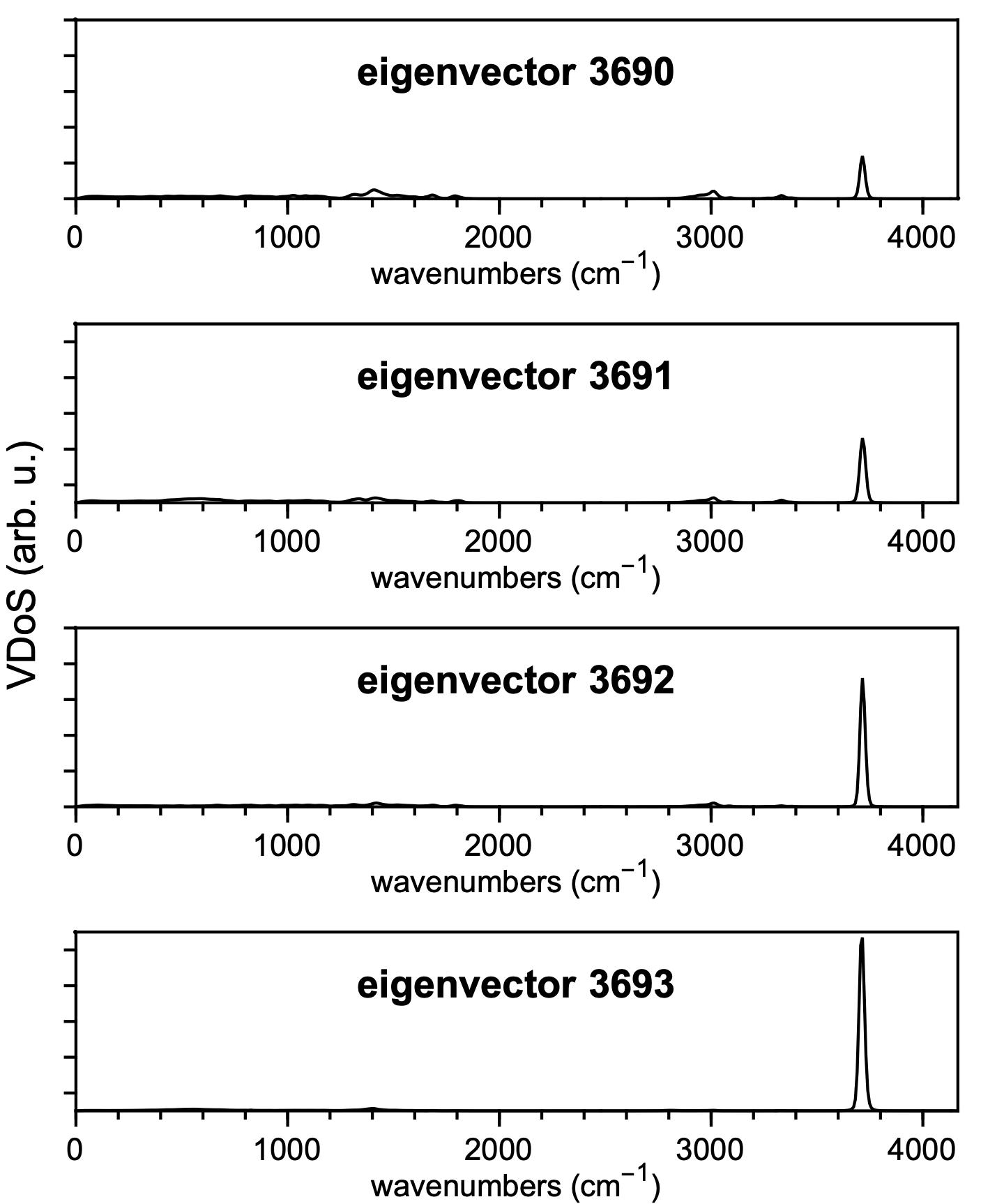}}
    \caption{
    Hot protein O-H vibrations. VDoS of single DOFs obtained after projection of the equilibrium trajectory on eigenvectors 3960-3963 computed from $\vc{C}(\tau=0)$ from the steady-state simulation with average temperatures of up to 540~K.
    }
    \label{f:last4}
\end{figure}

\newpage

\end{widetext}